\documentclass[prb,twocolumn,floats,superscriptaddress,showpacs]{revtex4}
\usepackage{amsmath}

\usepackage{epsfig}
\usepackage{graphicx}
\usepackage{dcolumn}
\usepackage{bm}

\def\gapp{\lower.35em\hbox{$\stackrel{\textstyle>}{\sim}$}}
\def\lapp{\lower.35em\hbox{$\stackrel{\textstyle<}{\sim}$}}
\begin{document}
\bibliographystyle{apsrev}
\title{Effect of the tetrahedral distortion on the electronic properties of iron-pnictides}
\author{M.J. Calder\'on}
\email{calderon@icmm.csic.es}
\affiliation{Instituto de Ciencia de Materiales de Madrid,
ICMM-CSIC, Cantoblanco, E-28049 Madrid (Spain)}
\author{B. Valenzuela}
\email{belenv@icmm.csic.es}
\affiliation{Instituto de Ciencia de Materiales de Madrid,
ICMM-CSIC, Cantoblanco, E-28049 Madrid (Spain)}
\affiliation{Departamento de la Materia Condensada, Universidad Aut\'onoma de Madrid,
Cantoblanco, E-28049 Madrid (Spain)}
\author{E. Bascones}
\email{leni@icmm.csic.es}
\affiliation{Instituto de Ciencia de Materiales de Madrid,
ICMM-CSIC, Cantoblanco, E-28049 Madrid (Spain)}
\date{\today}
\begin{abstract}

We study the dependence of the electronic structure of iron pnictides
on the angle formed by the arsenic-iron bonds. Within a Slater-Koster
tight binding model
which captures the correct symmetry properties of the bands,
we show that the density of states and the band
structure are sensitive to the distortion of the tetrahedral environment of the
iron atoms.
This sensitivity is extremely strong in a two-orbital (d$_{xz}$, d$_{yz}$) 
model due to the formation of a flat band around the Fermi
level.  Inclusion of the d$_{xy}$ orbital destroys the flat band while 
keeping a  considerable angle dependence in the band structure.
\end{abstract}
%
\maketitle

\begin{figure}
\leavevmode
\includegraphics[clip,width=0.3\textwidth]{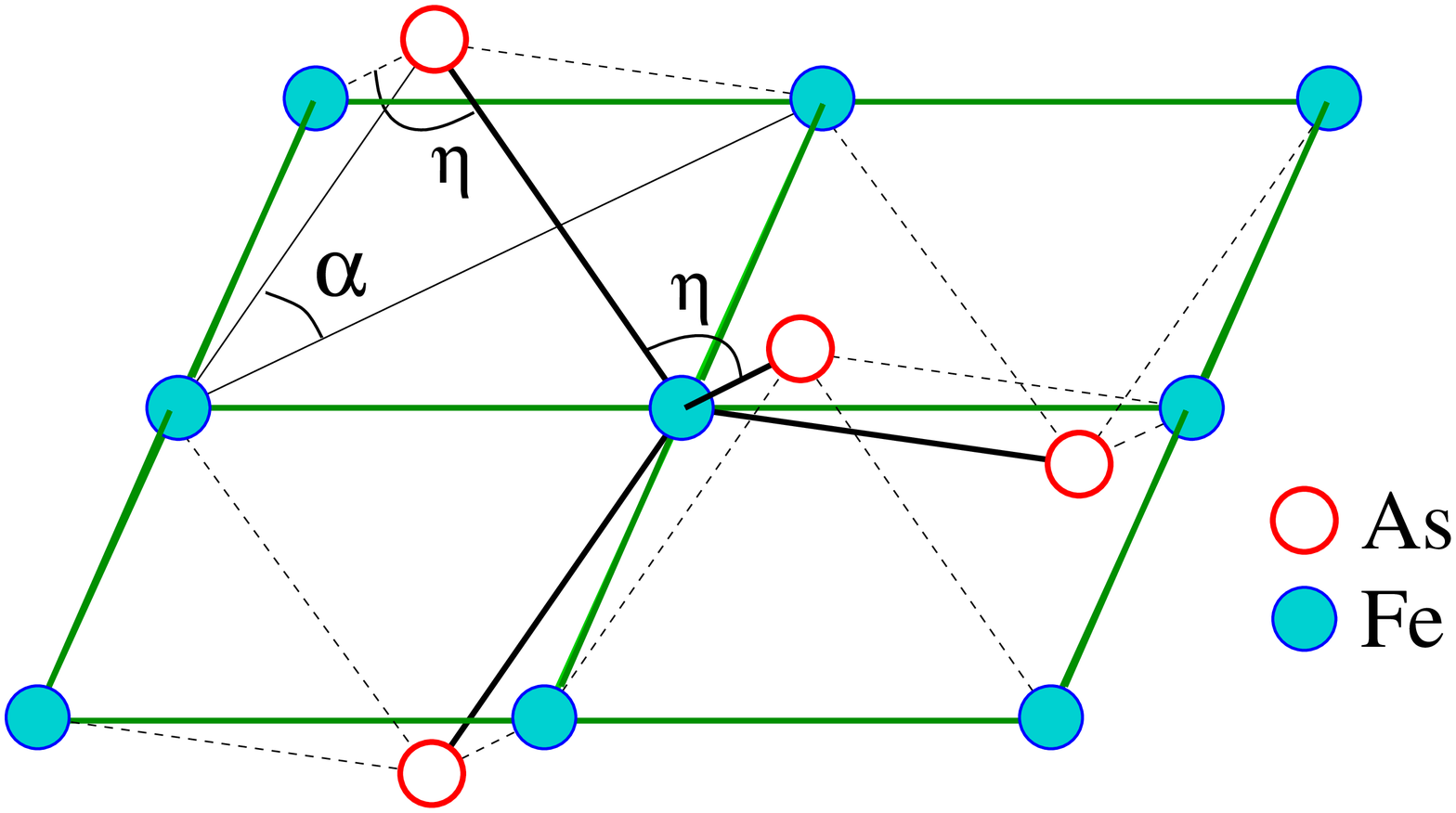}
\caption{Lattice structure of the FeAs layers with each Fe atom surrounded by
four As atoms in a distorted tetrahedral geometry. The Fe atoms form a square
lattice. At the center of each square, above and below in a checkerboard form,
lie the  As atoms. $\alpha$, the angle formed by the Fe-As bond and the
Fe-plane, is related to  the As$^{top}$-Fe-As$^{top}$ (or Fe-As-Fe, with the Fe atoms being next nearest neighbors) angle $\eta$ as
$2\alpha=180^o-\eta$. For instance, for a regular tetrahedron $\eta=109.47^o$ and $\alpha=35.26^o$.
}
 \label{fig:lattice}
\end{figure}

The recent discovery of high temperature superconductivity in iron
arsenides~\cite{kamihara08} has triggered off intense research 
in the condensed
matter community. The characteristic building blocks of these
materials are FeAs layers with each Fe atom surrounded by four
arsenic atoms in a distorted tetrahedral geometry. Fe atoms form a
square lattice, while As atoms located at the center of each square
are displaced above and below from the Fe-plane in a checkerboard
form (see Fig.~\ref{fig:lattice}). Undoped compounds are compensated
semimetals which undergo a structural and a magnetic transition.~\cite{cruz08,rotter08}
Upon chemical doping, the structural distortion and magnetic order
disappear and superconductivity emerges. Superconductivity has
been reported so far in three families: rare-earth 
oxyarsenides\cite{kamihara08} ReFeAsO, also denoted 1111 systems, with Re a
lanthanide atom, TFe$_2$As$_2$ with T an alkaline earth metal and two 
FeAs planes
per unit cell,~\cite{rotter08-2} called 122, and LiFeAs.~\cite{pitcher08} Iron arsenides can be
considered part of a larger family, the iron pnictides, with As 
substituted by an isovalent pnictogen element, as P.~\cite{kamihara06}

Hopping between Fe atoms via As is expected to give the dominant
contribution to the kinetic energy~\cite{cao08} and to the exchange
interaction.~\cite{yildirim08}
Therefore some dependence of the electronic properties on the angles formed by
the Fe-As bonds, i.e on the distortion of the As tetrahedron,
is expected. This dependence is of crucial importance for both
weak coupling models,~\cite{ma08,raghu08,mazin08-2} based on nesting 
properties, and strong 
coupling models,~\cite{yildirim08} based on superexchange interaction, aimed to explain the properties of these  
systems.
For each family of
iron-pnictides, the As tetrahedron presents different distortions.
It is almost regular~\cite{rotter08} in the 122 family with the four As-Fe-As angles
close to the ideal value 109.47$^o$, while
it is elongated  in LiFeAs,~\cite{pitcher08} with As atoms further from the Fe-plane than in the
regular tetrahedron (the As$^{top}$-Fe-As$^{top}$ angle
equals 102.8$^o$). Here As$^{top}$ refer to an As atom placed above
the Fe plane. In the 1111 family the tetrahedron is squashed and the
As$^{top}$-Fe-As$^{top}$ or  P$^{top}$-Fe-P$^{top}$ angle depends on specific composition, being
equal to 113.7$^o$ in LaFeAsO\cite{kamihara08} and 120.6$^o$ in LaFePO.~\cite{kamihara06} Differences in the electronic 
properties of these compounds have been attributed to the different 
distorsion of the tetrahedra.~\cite{mcqueen08} The As-Fe-As angle is also sensitive
to doping~\cite{lee08,zhao08,kasperkiewicz08,huang08,rotter08-3} and can be deeply modified under pressure.~\cite{kreyssig08} An
example of the latter is the collapsed tetragonal phase found in
CaFe$_2$As$_2$ under pressure.~\cite{kreyssig08} In Refs.~\onlinecite{lee08} and \onlinecite{zhao08} a
possible correlation between the superconducting critical
temperature and the As-Fe-As angle has been suggested with
deviations within a family from the regular tetrahedron being
detrimental for superconductivity. In Ba$_{1-x}$K$_x$Fe$_2$As$_2$ at optimal doping the tetrahedron changes from squashed (in underdoped)  to elongated (in overdoped).~\cite{rotter08-3} 
It has been argued~\cite{lee08,zhao08} that 
there is a correlation
between stronger interaction and higher critical temperature in iron
pnictides due to a narrower bandwidth for the regular tetrahedron. This 
argument seems to be supported by the lower critical temperatures and lack of 
structural distortion and magnetism in 
LiFeAs
and LaFePO.
An unusually large sensitivity 
of the iron moment~\cite{yin08} and the band
structure\cite{boeri08,mazin08-2,vildosola08} to the separation of the As 
atoms with respect to the
plane has been also found in Density Functional theory
calculations.

In this paper we analyze the
angle dependence of the band structure within a tight-binding model
in which hopping between Fe atoms is assumed to be mediated by As
and its magnitude is calculated within the Slater Koster framework.~\cite{slater54}
We show that the two-orbital model which only includes $d_{xz}$ and
$d_{yz}$ is extremely sensitive to changes in the angle formed by the As-Fe bonds. Such a sensitivity originates in the appearance of
flat bands in the $(0,0)-(\pm \pi,0)$ and $(0,0)-(0,\pm \pi)$
directions for the regular tetrahedron case. The corresponding peak
in the density of states is found at the Fermi level at half
filling. The flat bands disappear when the tetrahedron is distorted.
Within the present model this result is independent of any fitting
parameter. Inclusion of the d$_{xy}$ orbital modifies this picture
but maintains the dependence of the density of states and band structure
on the angle.

{\it Two-orbital model.} 
We construct a tight binding model to describe the band structure of the FeAs
layers around the Fermi level in the tetragonal phase.
We describe the distortion of the tetrahedron in terms of the angle
$\alpha$ formed by the Fe-As bond direction with the Fe-plane, see Fig.~\ref{fig:lattice}. The
advantage of using this angle, instead of the more common Fe-As-Fe or As-Fe-As
ones, is that it is uniquely defined. On the contrary, there are two different Fe-As-Fe angles (depending on whether the Fe atoms are nearest or next-nearest neighbors) and, for non-regular tetrahedra there are also two different As-Fe-As angles (between top As and between top and down As). The correspondence between
$\alpha$ and the Fe-As-Fe or As-Fe-As angles is straightforward (see figure
caption of Fig.~\ref{fig:lattice}).

According to first principles calculations several bands cross the Fermi level
and the density of states (DOS) at the Fermi level is dominated 
by Fe-d orbitals 
with little weight of pnictogen p-orbitals.~\cite{lebegue07,boeri08,cao08}
We work in the Fe-square lattice and take x and y axes
along the Fe-Fe bonds. Only the Fe-orbitals are included in the tight binding.
As atoms only enter in the model indirectly via the Fe-Fe hopping amplitudes.
Under these
assumptions the Hamiltonian is given by:
\begin{eqnarray}
H  &=&  \sum_{i,j,\beta,\gamma,\sigma} (t^x_{\beta,\gamma}
  c^\dagger_{i,j,\beta,\sigma}c_{i+1,j,\gamma,\sigma}+ t^y_{\beta,\gamma}
  c^\dagger_{i,j,\beta,\sigma}c_{i,j+1,\gamma,\sigma}  \nonumber \\
& +&  t'_{\beta,\gamma}c^\dagger_{i,j,\beta,\sigma}c_{i+1,j+1,\gamma,\sigma} +
t''_{\beta,\gamma}c^\dagger_{i,j,\beta,\sigma}c_{i+1,j-1,\gamma,\sigma}  \nonumber\\
&+& h.c. ) +  \sum_{i,j,\beta,\sigma}\epsilon_\beta
c^\dagger_{i,j,\beta,\sigma}c_{i,j,\beta,\sigma} -\mu
\label{eq:hamil}
\end{eqnarray}
Here $i,j$ and $\sigma$ label the sites in the 2D lattice and  the spin
respectively, while $\beta,\gamma$ refer to the Fe-d orbitals included
in the tight binding.
$\epsilon$ is the on-site energy, and $\mu$ is the chemical potential. 
We first restrict ourselves to two degenerate $d_{xz}$ and $d_{yz}$ orbitals, 
as in
other minimum models recently discussed.~\cite{dai08,daghofer08} 
Due to the degeneracy of d$_{xz}$ and d$_{yz}$ orbitals the onsite energies
$\epsilon_{xz}=\epsilon_{yz}$ only shift the bottom of the bands and for
simplicity we take them equal to zero.
The hopping amplitudes $t$ are calculated from orbital overlap integrals
within the Slater-Koster framework,~\cite{slater54} which
captures the
correct symmetry properties of the energy bands.
Direct Fe-Fe hopping is neglected, i.e.
we assume that all 
hopping takes
place via As atoms. Neglecting the
energy splitting between As p$_{x,y}$ and p$_z$ orbitals and to second order
in perturbation theory:
\begin{equation}
t^x_{xz,xz}=2\left[(t_{x,xz})^2-(t_{y,xz})^2-(t_{z,xz})^2\right]/(\epsilon_d-\epsilon_p)
\end{equation}
\begin{equation}
t^y_{xz,xz}=2\left[(t_{x,xz})^2-(t_{y,xz})^2+(t_{z,xz})^2\right]/(\epsilon_d-\epsilon_p)
\end{equation}
and $t^x_{yz,yz}=t^y_{xz,xz}$, $t^y_{yz,yz}=t^x_{xz,xz}$. 
Hopping to nearest neighbors between $xz$ and $yz$ vanishes. Next
nearest neighbor hopping amplitudes satisfy $t''_{xz,yz}=-t'_{xz,yz}$,
$t''_{xz,xz}=t'_{xz,xz}$, $t''_{yz,yz}=t'_{yz,yz}$ and
$t'_{yz,yz}=t'_{xz,xz}$ with
\begin{equation}
t^{'}_{xz,xz}=\left[(t_{x,xz})^2+(t_{y,xz})^2-(t_{z,xz})^2\right]/(\epsilon_d-\epsilon_p)
\end{equation}
\begin{equation}
t^{'}_{xz,yz}=\left[-t_{x,xz}t_{x,yz}-t_{y,xz}t_{y,yz}-t_{z,xz}t_{z,yz}\right]/(\epsilon_d-\epsilon_p)
\end{equation}
with  $\epsilon_p$ and $\epsilon_d$ the on-site energy of the pnictogen-p and
Fe-d orbital, and
$t_{\delta,\beta}$, with $\delta=x,y,x$ and $\beta=xz,yz$, the overlap between
As-p$_\delta$ and Fe-d$_{\beta}$ orbitals.
These expressions take into account that two As atoms mediate the hopping
between nearest neighbors, while only one As atom is involved in the
hopping between next nearest neighbors.
Overlap between As-p orbitals and Fe-d
orbitals is given in terms of two integrals, taken as disposable constants,
$pd_\sigma$ and $pd_\pi$, where as usual $\sigma$ and $\pi$ refer to the
component of angular momentum around the axis.~\cite{slater54}
The sign of $pd_\pi$ is taken opposite to that of
$pd_\sigma$.~\cite{harrison-book}
\begin{figure}
\leavevmode
\includegraphics[clip,width=0.5\textwidth]{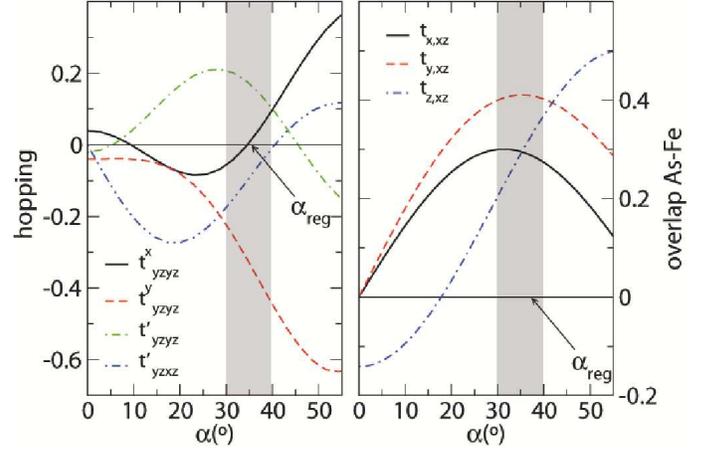}
\caption{Angle dependence of the Fe-Fe hopping parameters (left) and Fe-As
  overlap integrals (right) which enter in the
  two-orbital model. Fe-As overlap integrals and Fe-Fe hopping parameters
are in units of $pd_{\sigma}$ and 
$pd_{\sigma}^2/(\epsilon_d-\epsilon_p)$ respectively. $pd_{\pi}=-0.2$.
Shaded areas correspond to the experimentally relevant angles. 
}
 \label{fig:hoppings}
\end{figure}
The angle-dependence is evident in the Fe-As overlaps:
\begin{equation}
t_{x,xz}=\sin\alpha \left (\frac{\sqrt{3}}{2}\cos^2 \alpha (pd_\sigma) + \sin^2
\alpha (pd_\pi)\right)
\label{eq:txxz}
\end{equation}
\begin{equation}
t_{y,xz}={{1}\over{2}}\,\sin\alpha \,\cos^2 \alpha\left(\sqrt{3} (pd_\sigma)-2(pd_\pi)\right)
\end{equation}
\begin{equation}
t_{z,xz}=\frac{\cos\alpha}{\sqrt{2}}\left(\sqrt{3}\sin^2 \alpha
(pd_\sigma)+(1-2\sin^2\alpha) (pd_\pi)\right )
\label{eq:tzxz}
\end{equation}
and  $t_{x,yz}=t_{y,xz}$, $t_{y,yz}=t_{x,xz}$, and $t_{z,xz}=t_{z,yz}$. 
The angle dependence of the hopping parameters which enter in Eq.~(\ref{eq:hamil})
and of the Fe-As overlaps are shown in Fig.~\ref{fig:hoppings}. 
A strong 
dependence is seen, specially in the Fe-Fe hopping, for the angles found
experimentally (shaded area in Fig.~\ref{fig:hoppings}). 
In this figure Fe-Fe hopping parameters and Fe-As overlaps are given in 
units of $(pd_\sigma)^2/(\epsilon_p-\epsilon_d)$ and $pd_\sigma$ respectively. 
In these units
all the hopping amplitudes and overlap integrals depend on a unique free
parameter $pd_\pi$.

Diagonalization of the Hamiltonian results in two bands
\begin{equation}
E^\pm({\bf k})=\epsilon_+({\bf k})-\mu \pm \sqrt{\epsilon_-({\bf
k})^2+\epsilon_{xy}^2({\bf k})}
\label{eq:E+-}
\end{equation}
with
\begin{equation}
\epsilon_+({\bf k})= (t^x_{xz,xz}+t^y_{xz,xz})(\cos k_x+\cos k_y) +4 t'_{xz,xz}\cos k_x \cos k_y
\end{equation}
\begin{equation}
\epsilon_-({\bf k})=(t^x_{xz,xz}-t^y_{xz,xz})(\cos k_x-\cos k_y)
\end{equation}
\begin{equation}
\epsilon_{xy}({\bf k})=4 t'_{yz,xz}\sin k_x \sin k_y\nonumber
\end{equation}
\begin{figure}
\leavevmode
\includegraphics[clip,width=0.5\textwidth]{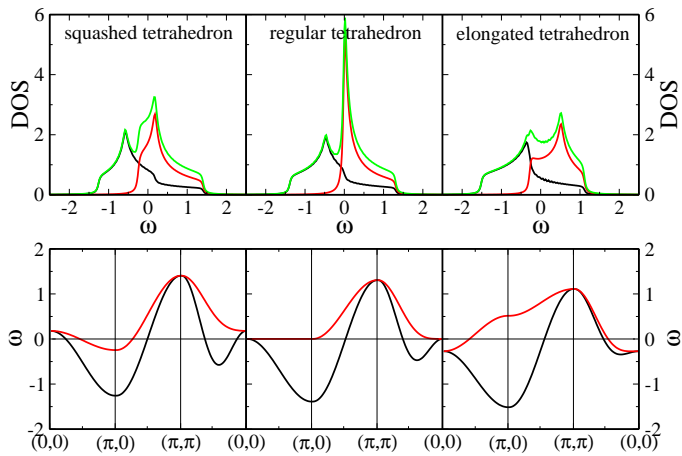}
\caption{Top (bottom) figures: Density of states (band structure) for the
  two-orbital model
  corresponding to  the squashed (left), regular (middle) and elongated
  (right) tetrahedron. The corresponding angles are $\alpha=32.71^o$, $35.26^o$ and $39.13^o$, respectively.
  In the top figures the green (light) lines show the total density of
  states. Black and red lines give the density of states corresponding to the
  $E^-$ and $E^+$ bands (Eq.~\ref{eq:E+-}) respectively. The density of states
  is smoothed by a lorentzian of width $0.02$. Energies are in units of
  $pd_{\sigma}^2/(\epsilon_d-\epsilon_p)$. 
}
\label{fig:dos-bands}
\end{figure}
The density of states (DOS) for a squashed ($\alpha=32.71^o$), a regular ($\alpha=35.26^o$) and an
elongated tetrahedron ($\alpha=39.13^o$), corresponding to As$^{top}$-Fe-As$^{top}$ angles of
$114.58^o$, $109.47^o$ and $101.74^o$ respectively, and $pd_\pi= -0.2$ is plotted in
Fig.~\ref{fig:dos-bands} (top panels).
In this figure the chemical potential is at $\omega=0$, assumming half-filling.
Two peaks are observed in the DOS for the three angles. The DOS of each of
the two bands
$E^\pm$ (Eq.~\ref{eq:E+-}) is also plotted in these figures. It can be seen that each of the
peaks comes from one of the two bands. The height of the high energy peak is very strongly sensitive to $\alpha$. In particular,
the intensity of
this peak is much higher for a regular tetrahedron. Inspection of the band structure in the lower plots of
Fig.~\ref{fig:dos-bands} reveals the existence of a flat band along $(0,0)-(\pi,0)$ in this case. By symmetry,
the band is flat also along the $(0,0)-(0,\pm \pi)$ and $(0,0)-(-\pi,0)$
directions. The flat bands in the regular tetrahedron geometry ($\alpha=35.26^o$) appear for any value of $pd_\sigma$ and $pd_\pi$. From the expressions for
the hopping amplitudes given above, for this $\alpha$ it can be shown that $t^x_{xz,xz}+2 t'_{xz,xz}=0$,
which cancels the $k_x$
dependence of the band structure for $k_y=0$ (alternatively, it cancels the
$k_y$-dispersion for $k_x=0$). 
This equality for the regular tetrahedron comes from $t_{x,xz}=t_{z,xz}$, see
Eqs.~(\ref{eq:txxz}) and (\ref{eq:tzxz}) and right panel in Fig. 2.
The topology of the energy contours and, therefore, that of the Fermi surface, is also strongly sensitive to $\alpha$ as
shown in Fig.~\ref{fig:contour} for the same parameters used in Fig.~\ref{fig:dos-bands}.
\begin{figure}
\leavevmode
\includegraphics[clip,width=0.45\textwidth]{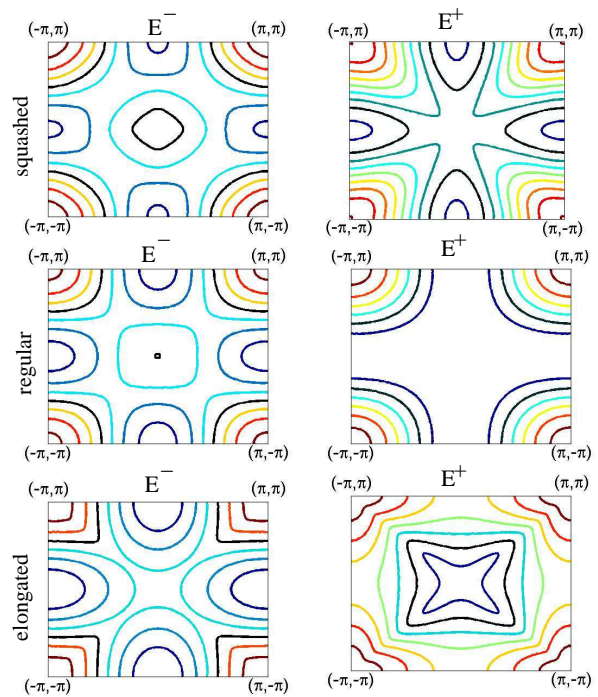}
\caption{Energy contour plots for the squashed (top), regular (middle) and
  elongated (bottom) tetrahedral geometries for the same parameters as in
  Fig.~\ref{fig:dos-bands}. Left figures correspond to $E^-$ and right figures to $E^+$, see Eq.~\ref{eq:E+-}. The Fermi surface are in black (darker lines). The topology of the
Fermi surface changes with the angle.}
 \label{fig:contour}
\end{figure}

Due to the change in the energy contour (and the Fermi surface) topology, and the very high peak in the DOS, if
interactions were included,
the magnetic and superconducting properties of this two-orbital model  
would be extremely sensitive to the angle $\alpha$.
This is particularly evident in a picture which ascribes the
magnetic transition to a spin density wave due to nesting.
We show now that including a third orbital maintains some angle dependence in 
the electronic properties although when the hybridization with the third
orbital is strong the conditions for the flat band are not fulfilled.
 
\begin{figure}
\leavevmode
\includegraphics[clip,width=0.4\textwidth]{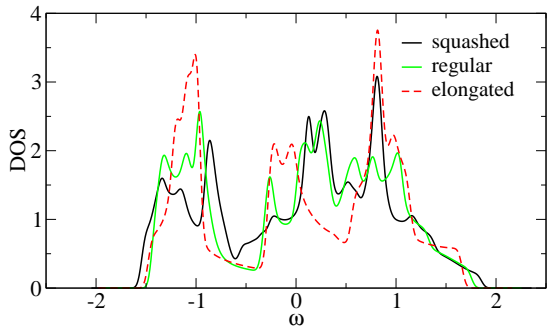}
\caption{Density of states for the three orbital model corresponding to the
  squashed, regular and elongated for the same angles
  as in Fig.~\ref{fig:dos-bands}, $pd_\pi=-0.2$ and $\epsilon_{xy}-\epsilon_{xz,yz}=0.15$. The density of states
  is smoothed by a gaussian of width $0.03$. Energies are in units of
  $pd_{\sigma}^2/(\epsilon_d-\epsilon_p)$. 
}
 \label{fig:3orbs}
\end{figure}

{\it Three orbital model}.
According to LDA~\cite{lebegue07,boeri08,cao08} the d$_{xy}$ orbital contributes to the Fermi surface and it
is expected to be relevant in any description of iron pnictides.~\cite{patricklee08} Hamiltonian
(\ref{eq:hamil}) can be straightforwardly generalized to include the d$_{xy}$ or any of the e$_g$ orbitals.~\cite{slater54}
In the case of the regular tetrahedron, including only the crystal field due
to the arsenic environment, the three t$_{2g}$ orbitals (d$_{xy}$, d$_{yz}$, and d$_{xz}$) are degenerate.~\cite{boeri08} The inclusion of the crystal field produced by the Fe environment and/or the distortion of the
tetrahedra breaks this degeneracy. As the As atom approaches the Fe-plane the
d$_{xy}$ orbital becomes higher in energy as compared to the (still degenerate)
d$_{xz}$ and d$_{yz}$ orbitals.
The level splitting $\epsilon_{xy}-\epsilon_{xz,yz}$ introduces a new
parameter in the calculation.
When both Fe and As crystal field effects
are included $\epsilon_{xy}-\epsilon_{xz,yz}$ is slightly larger
for the squashed tetrahedron than
for the regular or elongated one. Expansion of hopping parameters to second
order in perturbation theory also contributes to the renormalization of the on-site
energies $\epsilon_{xy}$ and $\epsilon_{xz,yz}$.

For large level splittings $\epsilon_{xy}-\epsilon_{xz,yz} \gtrsim W$, with
$W$ the bandwidth of the two-orbital model, hybridization between $d_{xy}$
and the bands $E^+$ and $E^-$ of the two-orbital model is weak and the picture described 
above remains.
However, in real materials the level splitting is expected to be smaller.~\cite{boeri08}
Fig.~\ref{fig:3orbs} shows the DOS of the three orbital model for the same
geometries as in Fig.~\ref{fig:dos-bands} and with
$\epsilon_{xy}-\epsilon_{xz,yz}=0.15$, and $pd_\pi=-0.2$. 
For simplicity, we neglect the angle-dependence of the level splitting. Fig.~\ref{fig:3orbs} reveals that a clear dependence of the DOS on the angle $\alpha$ 
still prevails within the
three-orbital model. Note that the Fermi level is expected to be around the center of the band where the
effect of changing the angle is very strong.~\cite{shorikov08} 
However, the origin of the dependence of the DOS on the angle $\alpha$ is qualitatively different to the one found in the two-orbital model: 
Due to the hybridization with
the $d_{xy}$ orbital  the
flat band that appeared in the regular tetrahedron geometry within the
two-orbital model is partially destroyed in the three-band model.

In conclusion, we have analyzed the angle dependence of the band structure
within a Slater-Koster based tight binding model for iron pnictides. In both
two and three orbital models the density of states  and therefore the Fermi
surface are strongly sensitive to
changes in the angle formed between the Fe-As bonds.  This result is
relevant for both weak coupling theories based on the nesting mechanism,
which is very sensitive to the shape of the Fermi surface, and for strong
coupling theories, since superexchange also depends on orbital overlaps. In the two-orbital model
a flat band is formed when the As-tetrahedron is regular what results in
a strong peak in the density of states. This result is
robust against changes in the fitting parameters. Hybridization with a third
orbital modifies this picture although the dependence on the angle remains, 
in particular for low energies. In our description, the bandwidth does not seem 
to depend very strongly on the angle, though it seems to be slightly narrower for a
more elongated tetrahedron.

We thank S. Biermann, A. Cano, A. Cortijo, P. Monod, R. Roldan, A.
Santander and  V. Vildosola for useful conversations. E.B. thanks
the hospitality of ESPCI-Paris where part of this work was done. We
acknowledge funding from Ministerio de Ciencia e Innovaci\'on
through Grants No. FIS2005-05478-C02-01 and MAT2006-03741, Ram\'on y
Cajal contracts and Jos\'e Castillejo program, and from
Consejer\'{\i}a de Educaci\'on de la Comunidad Aut\'onoma de Madrid
and CSIC through Grant No. CG07-CSIC/ESP-2323.

\end{document}